\documentclass[lettersize,journal]{IEEEtran}
\usepackage{amsmath,amsfonts}
\usepackage{algorithmic}
\usepackage{algorithm}
\usepackage{array}
\usepackage[caption=false,font=normalsize,labelfont=sf,textfont=sf]{subfig}
\usepackage{textcomp}
\usepackage{stfloats}
\usepackage{url}
\usepackage{verbatim}
\usepackage{graphicx}
\usepackage{cite}
\usepackage{multirow}
\usepackage{bbm}
\usepackage[colorlinks, linkcolor=cyan, anchorcolor=cyan, citecolor=cyan]{hyperref}
\usepackage{booktabs}
\newcommand{\etal}{{\textit{et al.}}}
\newcommand{\ie}{{\textit{i.e.}}, }
\newcommand{\eg}{{\textit{e.g.}}, }

\begin{document}
\title{PointNu-Net: Keypoint-assisted Convolutional Neural Network for Simultaneous Multi-tissue Histology Nuclei Segmentation and Classification}
\author{Kai Yao, Kaizhu Huang, Jie Sun  and  Amir Hussain
\thanks{Kai Yao is with both University of Liverpool and School of Advanced Technology, Xi'an Jiaotong-Liverpool University.}
\thanks{Kaizhu Huang is with Duke Kunshan University, Suzhou, Kunshan, 215316, China.}
\thanks{Jie Sun is with  School of Advanced Technology, Xi'an Jiaotong-Liverpool University, Suzhou, Jiangsu, 215000, China.}
\thanks{Amir Hussain is with School of Computing, Edinburgh Napier University, Edinburgh, EH11 4BN, the United Kingdom. } 
\thanks{Correspondence: 
kaizhu.huang@dukekunshan.edu.cn and jie.sun@xjtlu.edu.cn.}
\thanks{© 2023 IEEE.  Personal use of this material is permitted.  Permission from IEEE must be obtained for all other uses, in any current or future media, including reprinting/republishing this material for advertising or promotional purposes, creating new collective works, for resale or redistribution to servers or lists, or reuse of any copyrighted component of this work in other works}
}

\maketitle

\begin{abstract}
Automatic nuclei segmentation and classification play a vital role in digital pathology. However, previous works are mostly built on data with limited diversity and small sizes, making the results questionable or misleading in actual downstream tasks. In this paper, we aim to build a reliable and robust method capable of dealing with data from the ‘the clinical wild’. Specifically, we study and design a new method to simultaneously detect, segment, and classify nuclei from Haematoxylin and Eosin (H\&E) stained histopathology data, and evaluate our approach using the recent largest dataset: PanNuke. We address the detection and classification of each nuclei as a novel semantic keypoint estimation problem to determine the center point of each nuclei. Next, the corresponding class-agnostic masks for nuclei center points are obtained using dynamic instance segmentation. Meanwhile, we proposed a novel Joint Pyramid Fusion Module (JPFM) to model the cross-scale dependencies, thus enhancing the local feature for better nuclei detection and classification. By decoupling two simultaneous challenging tasks and taking advantage of JPFM, our method can benefit from class-aware detection and class-agnostic segmentation, thus leading to a significant performance boost. We demonstrate the superior performance of our proposed approach for nuclei segmentation and classification across 19 different tissue types, delivering new benchmark results. Code available: \href{https://github.com/Kaiseem/PointNu-Net}{https://github.com/Kaiseem/PointNu-Net}
\end{abstract}



\section{Introduction}
\label{sec:introduction}
\IEEEPARstart{A}{nalysing} digital pathology images can provide valuable diagnostic and prognostic cancer indicators to prompt precision medicine. These images usually contain tens of thousands of nuclei with various types. Their distribution and appearance are essential markers for cancer study. Automated analysis of these images can quantify biomarkers and histopathological features, reduce the workload of pathologists, and standardise clinical practices. This creates a huge demand for high-throughput nuclei detection, segmentation, and classification~\cite{chow2012nuclear}.
For instance, nuclear morphometric and appearance features can deliver massive tumor information for clinical prediction, e.g., diagnosis of disease grade and type~\cite{diagnosis}, disease prognosis and survival prediction~\cite{prognosis}, and  cancer metastases detection~\cite{metastases}. Moreover, classifying nuclei instead of patches in tissue may improve the prediction of recurrence and outcome  in some cases, e.g., in colon cancer~\cite{jiang2020machine}. 

\begin{figure}[t]
\centering
\includegraphics[width=0.9\columnwidth]{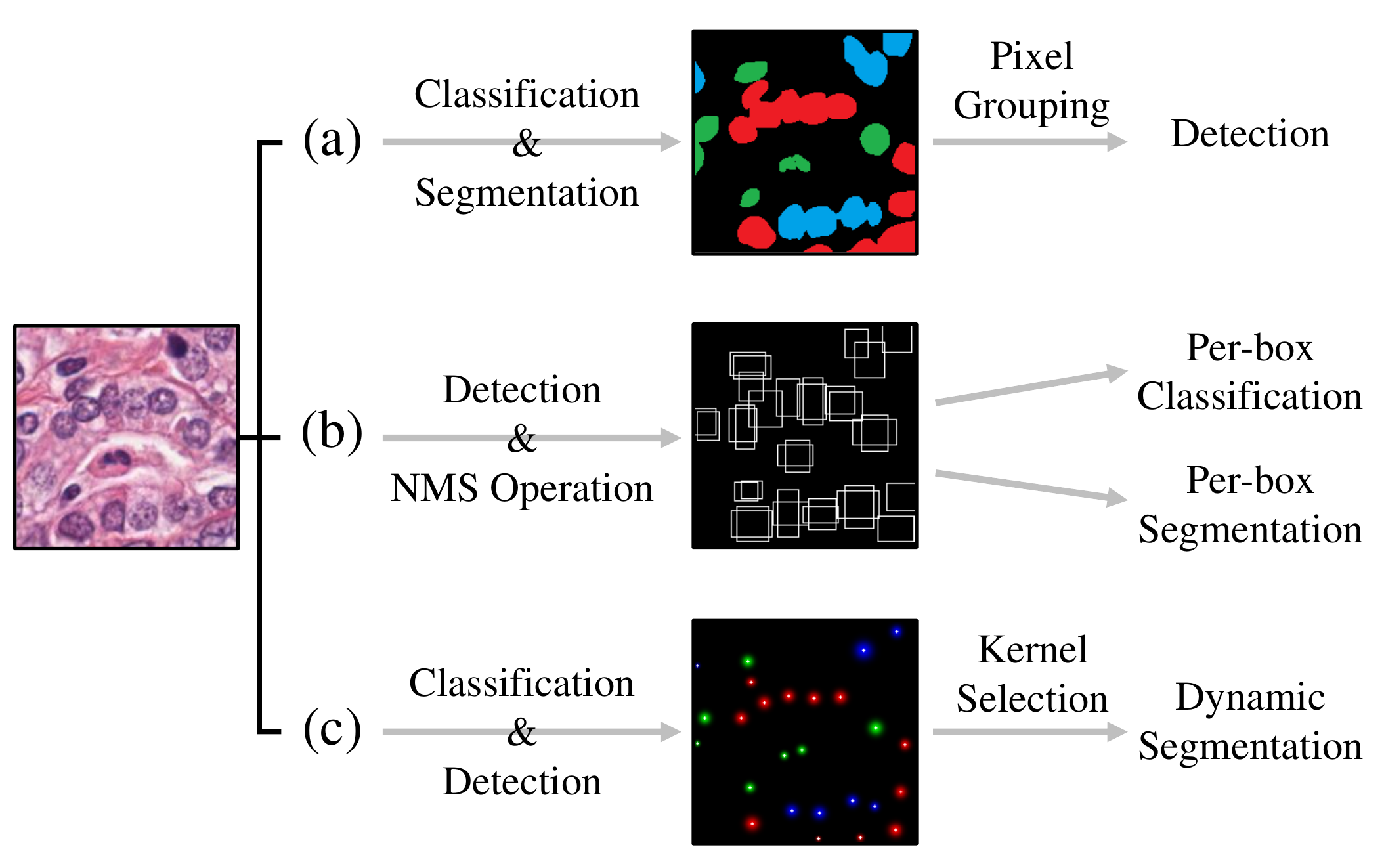}
\caption{\textbf{Difference among (a) bottom-up methods, (b) top-down methods, and (c) our method.} Bottom-up methods first predict semantic segmentation masks and then group the pixels into instances. Top-down methods first generate dense region proposals and conduct non-maximum suppression to obtain region-of-interests (ROIs), then classify and segment each ROI independently. Different from previous works, our method detects and classifies each nuclei via the estimation of semantic keypoint heatmap, and then class-agnostic masks are generated dynamically for each detected center point of nuclei.}
\label{fig:diff}
\end{figure}

Convolutional neural network (CNN) has exhibited superior performance in many applications of computer vision~\cite{ETCI1,ETCI2}. It has been introduced as a novel tool to process digital pathology images and identify morphological patterns in computational pathology~\cite{bera2019artificial,song2018contour,liu2021panoptic}. In general, on histology nuclei segmentation and classification, the present CNN-based approaches can be categorized into either bottom-up or top-down methods. The bottom-up structure is adopted by most existing methods \cite{kumar,dist,hovernet,cianet,tripleunet} which first generate high-resolution semantic segmentation masks and then group the pixels into an arbitrary number of object instances, as shown in \autoref{fig:diff}~(a). Relying on complicated pixel grouping post-processing to extract object instances, the performance of bottom-up approaches is highly dependent on segmentation results and grouping methods. Meanwhile, top-down methods, e.g., Mask-RCNN~\cite{maskrcnn}, first locate class-agnostic objects in prior bounding boxes to generate region proposals and then segment and classify object instances within region-of-interests (ROIs) (see \autoref{fig:diff}~(b)). Though Mask-RCNN can well separate touching nuclei, it can only segment the instances within bounding boxes with a very low resolution (i.e. $28\times28$). Furthermore, if bounding boxes are predicted less accurately and smaller than the actual instances, top-down methods may lead to poor segmentation. 

In addition to the drawbacks mentioned above, more seriously, most of these present works are typically built and evaluated on small datasets with limited diversity~\cite{kumar,hovernet}, mainly because the annotation of digital pathology images requires a large amount of time and effort from domain experts. 
As a result, the validity of the above-proposed methods may be  questionable~\cite{quest} and may cause misleading results in downstream analysis. In fact, there are a number of even more difficult challenges in 'the clinical wild'. First, the appearance of nuclei is obviously affected by manual operations such as slightly out-of-focus, imbalanced H\&E staining and blurred boundaries; folding tissue and hollow structure may also cause inconsistency of digital pathology images. Second, significant morphological changes can be observed across the real images. The nuclei of a single cell type can show varied shapes and sizes due to different diseases. Third, the density of the malignant cells in neoplastic tissues is usually much higher than the cell density in normal tissues, requiring that the analysis methods should have good generalization capabilities to process mutually occluded and overlapping nuclei under such extremely high cellularity and nuclei pleomorphism scenarios. All these challenges were not properly manifested in the current small data and are consequently not addressed effectively for nuclei images in 'the clinical wild'.

\begin{figure*}[!t]
\centering
\includegraphics[width=1.6\columnwidth]{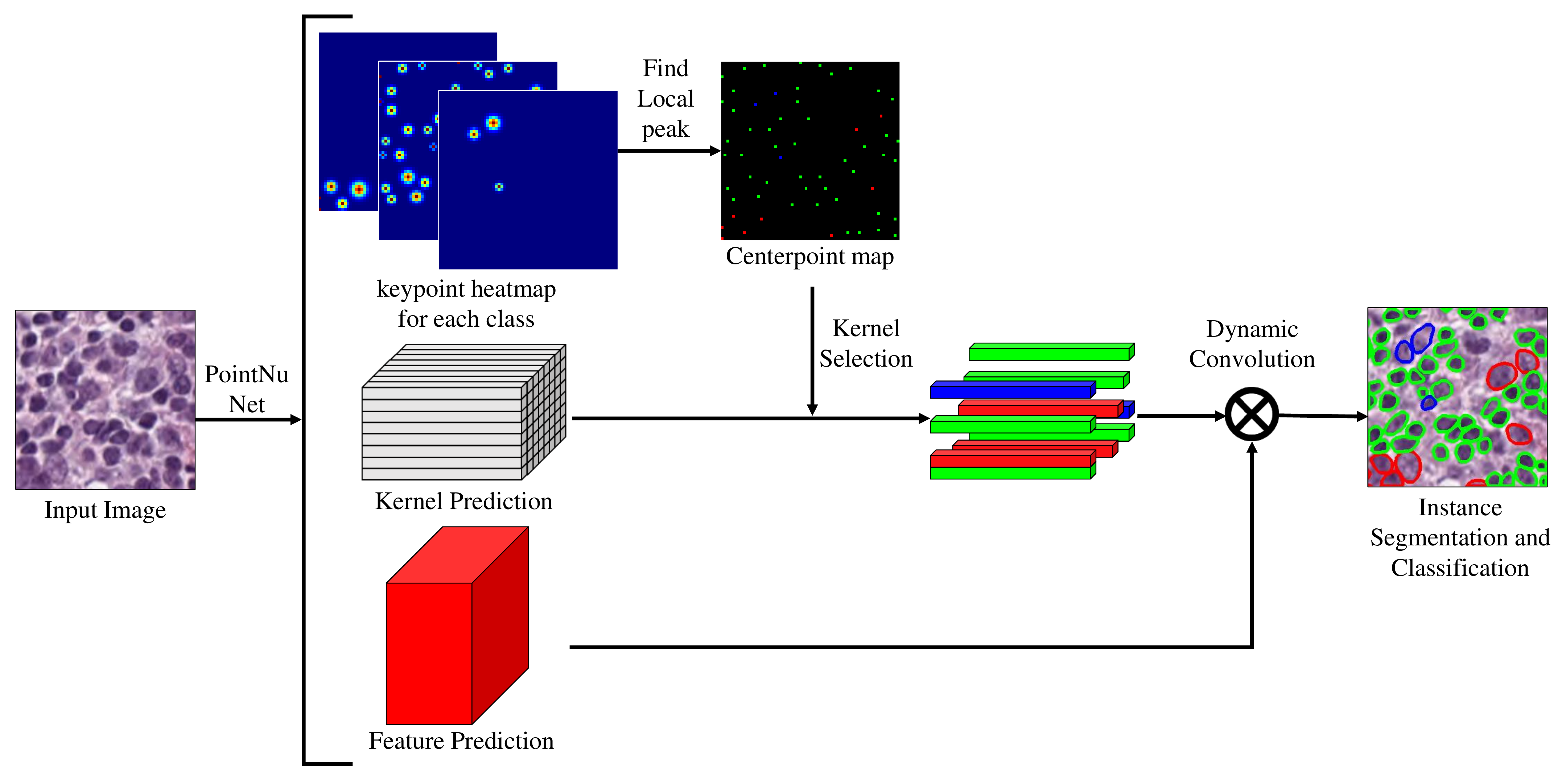}
\caption{\textbf{Overview of PointNu-net.} PointNu-net first produces semantic keypoint heatmaps to locate centers of nuclei and then generates high-resolution class-agnostic masks corresponding to the center position via dynamic convolution. Each color represents one class.}
\label{fig:dy}
\end{figure*}

Aiming to alleviate the problems in bottom-up and top-down methods as well as  building a reliable and robust method capable of dealing with challenges from the ‘the clinical wild’, in this paper, we study and design a new method to simultaneously detect, segment, and classify the nuclei from Haematoxylin and Eosin (H\&E) stained histopathology data. Specifically, different from both the bottom-up and top-down methods, we address the nuclei segmentation and classification problem from a new perspective and propose a novel keypoint-aware network that directly outputs instance segmentation and classification in histology images, as shown in \autoref{fig:diff}~(c). We develop a novel framework to segment and classify touching or overlapping nuclei by considering the problem as two simultaneous prediction problems. The network exploits keypoint heatmap regression to predict the center \textbf{point} of each \textbf{nu}clei (which inspires the model name \textbf{PointNu-Net}) that can detect and classify nuclei effectively; for each detected center point of instance, a high-resolution binary segmentation mask is then predicted using dynamic convolution.  

To our best knowledge, this is a brand new idea that was first applied to nuclei segmentation and classification. With the novel design, our method enjoys many benefits: compared with bottom-up methods,  better detection results  can be obtained by keypoint estimation without complex pixel grouping post-processing; compared with top-down methods, we utilize center points to represent nuclei instead of bounding boxes which can separate touching objects more precisely, and output high-resolution masks directly via dynamic instance segmentation. Therefore, overlapping and clustered nuclei can be separated as keypoints, leading to much better nuclei detection performance. In addition, our method can directly output instance segmentation and classification without any post-processing, enabling a very fast inference. 

To validate our method, we first present comparative results on various datasets including two famous multi-tissue histology image datasets. Our novel method demonstrates state-of-the-art performance compared to other recently proposed methods on nuclei segmentation and classification tasks. Importantly, we also conduct evaluations on the PanNuke data, the recently released largest pan-cancer histology dataset containing over 20,000 WSIs for nuclei segmentation and classification~\cite{gamper2020pannuke}, which is statistically similar to ‘the clinical wild’ and seldom investigated by previous methods. In such 'clinical wild', our proposed method leads to more robust and promising performance, which is substantially and consistently higher than all the comparison models for 19 tissue types.

Furthermore, our empirical findings indicate that high-resolution feature extractors are more efficacious in facilitating nuclei segmentation. Based on this discovery, we devise the Joint Pyramid Fusion Module (JPFM) to further optimize the utilization of high-resolution features. Unlike other neck structures such as Feature Pyramid Network (FPN)~\cite{fpn} and Atrous Spatial Pyramid Pooling (ASPP)~\cite{chen2017deeplab}, which tend to capture long-range contextual information, JPFM is designed to reinforce the extraction of local features by establishing task-specific dependencies across scales.

The main contributions of this work are three-fold: 
\begin{itemize}
\setlength{\itemsep}{0pt}
\setlength{\parsep}{0pt}
\setlength{\parskip}{0pt}
\item We propose a novel keypoint-aware network named PointNu-Net for nuclei segmentation and classification, where the keypoint heatmap is used to detect and classify the center point of each nuclei and the dynamic convolution is used to generate instance segmentation mask.
\item We propose JPFM to enhance the local feature by modeling cross-scale dependencies for better performance on detecting and classifying nuclei.
\item We evaluate our proposed method on one dataset which is statistically similar to ‘the clinical wild’ and with minimal selection bias. Experimental results show that the proposed PointNu-Net achieves state-of-the-art nuclei detection, segmentation and classification performance, as compared to recently published methods.
\end{itemize}
\section{Related Work}
In this section we give an overview of the related work. We first review bottom-up and top-down  nuclei segmentation and classification methods. Then we review some recently-developed dynamic instance segmentation methods as well as keypoint-based object detection approaches.
\subsection{Bottom-up Nuclei Segmentation And Classification}
Most histology nuclei segmentation and classification approaches are designed bottom-up.  Kumar \etal\cite{kumar} predicted three-class segmentation masks containing inside, outside, and edges of nuclei. Then pixels were grouped into instances using inside pixel seeded region growth. DIST~\cite{dist} formulated the inside nuclei segmentation problem as a regression of distance maps. CIA-Net~\cite{cianet} utilized additional contour supervision to obtain segmentation  with more accurate edges of nuclei. Triple U-Net~\cite{tripleunet} designed the parallel feature aggregation network to fuse features from Hematoxylin and RGB images progressively. Thus it learned a more precise nuclei boundaries. Hover-Net~\cite{hovernet} was the first to achieve simultaneous nuclei segmentation and classification. It outputs the instance center by regressing vertical and horizontal distances of nuclear pixels to their centers of mass and then predicted an additional nuclei type map to reach nuclei classification. 

Although these approaches achieved encouraging results, the detection performance of the bottom-up methods highly depends on the segmentation results and pixel grouping methods, thus  limiting their generalization  in complex scenarios, \eg in 'the clinical wild'. In contrast, we propose to regress a keypoint heatmap independently from nuclei segmentation. Thus there is no need of pixel grouping to separate nuclei, which enables  better generalization in complex scenarios.

\begin{figure*}[t]
\centering
\includegraphics[width=1.5\columnwidth]{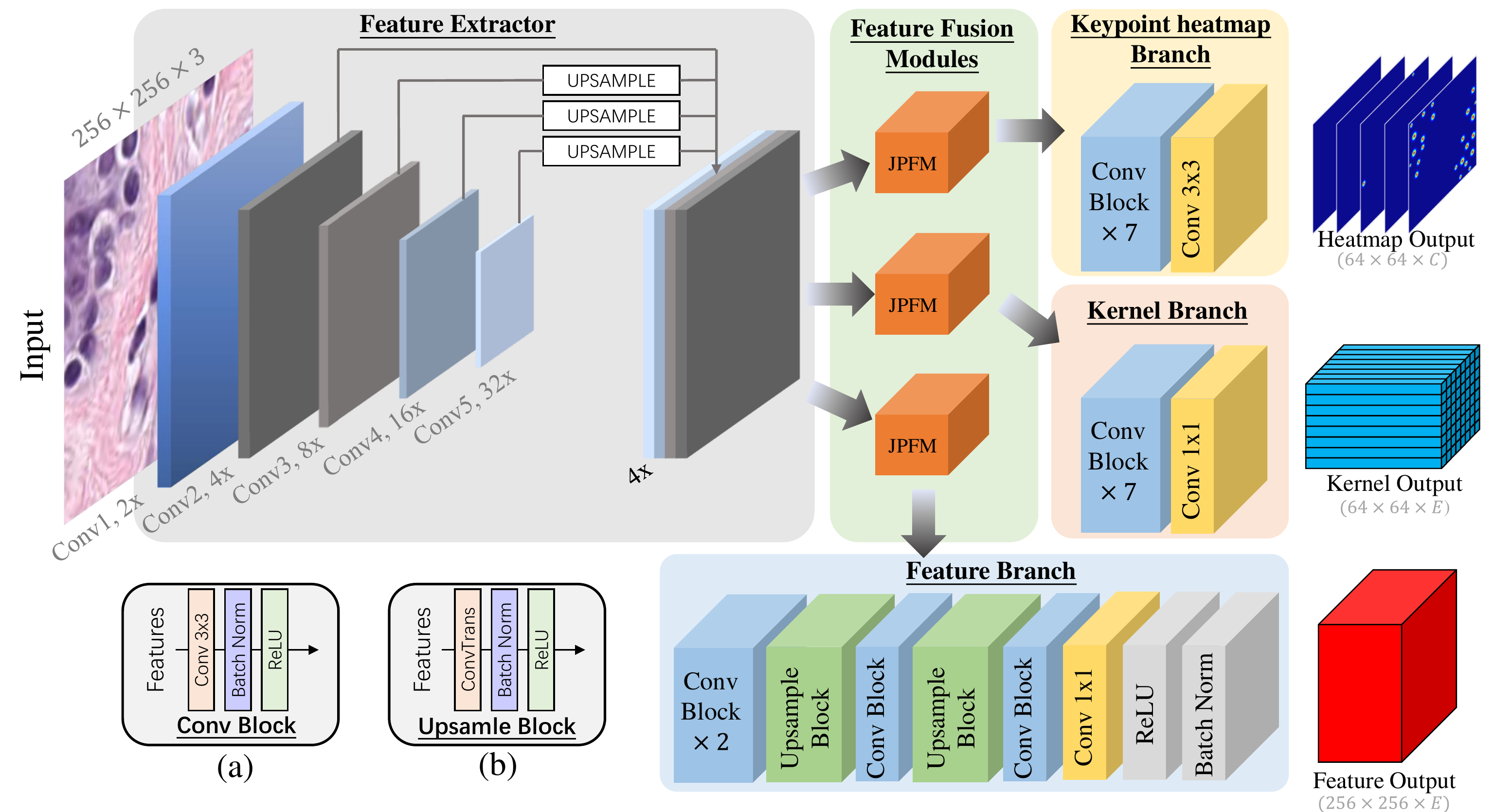}
\caption{\textbf{Network architecture.} (a) Conv block, (b) Upsample Block. The extracted multi-scale features are up-sampled to the same scale and concatenated before delivering to each branch. Keypoint heatmap branch predicts the center point of each nuclei as the detection and classification results. Then the outputs from kernel branch and feature branch are taken together and dynamic convolution is operated to get the instance segmentation results. }
\label{fig:arc}
\end{figure*}

\subsection{Top-down Nuclei Segmentation And Classification}
Top-down methods, \eg Mask-RCNN~\cite{maskrcnn}, have made rapid progress in natural image segmentation. Despite their potential in handling touching and clustered nuclei, they were seldom used in nuclei segmentation and classification. One major limitation of top-down  methods is the difficulty in segmenting precise nuclei boundaries. This may be caused by several reasons. 1). Since the prior anchors were usually designed  densely, one candidate region proposal may correspond to many overlapped bounding boxes. Thus, Non-Maximum Suppression (NMS) is often performed to filter the bounding boxes with low scores. This may result in poor segmentation when the bounding boxes cannot cover the whole nuclei. 2). Region proposal methods often have a fixed resolution of output segmentation masks (\eg $28 \times 28$). The predicted masks are then resampled to the corresponding bounding boxes' sizes, which may also introduce quantization errors.

On the contrary, since our method utilizes the peak of keypoint heatmap to represent object instances, one candidate proposal tends to be one point only. Thus, NMS can be optional but not necessary. In addition, our method predicts high-resolution masks for all instances via dynamic convolution, where the predicted masks have the same size as the input images. This can boost the performance not only in detection but also in segmentation compared with top-down methods.

\subsection{Dynamic Instance Segmentation}

Compared with traditional convolution, which utilizes a fixed and input-independent filter, dynamic convolution engages flexible kernels that are dynamically generated by another network. This idea has been explored previously in many areas. For instance,  Deformable Convolutional Networks~\cite{zhu2019deformable} learn the feature locations to be convoluted conditioned on the input features, enabling a sparse space sampling to capture global dependency; AdaIN~\cite{adain} applies input-dependent affine parameters to instance normalization layers for style injection within the image-to-image translation. Recent works investigate location-adaptive dynamic convolution in the instance segmentation task. The spatial positions of the feature map are treated as point proposals so as to distinguish instances. Similar to AdaIN, AdaptIS~\cite{Sofiiuk_2019_ICCV} generates affine parameters adaptively with the point proposal features. It then scales and shifts the normalized global features to predict instance masks. CondInst~\cite{condinst} generates the convolutional kernels, and assigns relative coordinates to the final feature to obtain the final prediction for each instance. SOLOv2~\cite{solov2} engages absolute positions to generate unified features in one shot for all predicted dynamic kernels. Similar to the previous works which are designed for natural images, we utilize dynamic convolution for nuclei instance segmentation. The difference is that, we perform keypoint heatmap estimation to boost the detection of nuclei (as detailed in \autoref{sec:abla1}), instead of recognizing objects through centerness.

\subsection{Keypoint-based Object Detection}
Unlike anchor-based detectors which regress objects by fitting the anchor boxes, keypoint-based object detectors directly regress object locations by utilizing features at certain pre-defined object keypoint. There are mainly two types of keypoint-based object detectors, \ie group-based  and group free, in the literature. Group-based detectors predict multiple keypoints for each object and group them to get bounding boxes. CornerNet~\cite{Cornernet} detected top-left and bottom-right corners of the object and then matched corners of the same object by computing the distance of points in the feature space. Duan~\etal\cite{duan2019centernet} added a center detection branch to improve the performance by center point validation. RepPoints~\cite{reppoints} took advantage of deformable convolutional networks to get sets of points to represent objects. CentripetalNet~\cite{centripetalnet} improved CornerNet by predicting the centripetal shift in order to pair corner keypoints from the same object. On the other hand, group-free detectors directly predicted the center keypoints of objects so as to avoid the complex grouping process. CenterNet~\cite{CenterNet} located objects by the center points and regressed the corresponding size. SaccadeNet~\cite{saccadenet} improved center point detection by extracting more informative corner features during training.

Since cell nuclei tend to cluster with high density, we utilize the group-free keypoint-based object detector to detect the center point of nuclei efficiently and effectively. Different from the above-mentioned methods, our proposed method decouples instance segmentation and classification into class-aware keypoint regression and class-agnostic mask prediction. This allows us to directly predict semantic instance masks in one shot without the need of pixel grouping post-processing, thus enabling both more precise and faster detection. To the best of our knowledge, our proposed method is the first approach that utilizes keypoint-based detection into nuclei instance segmentation.  

\begin{figure*}[t]
\centering
\includegraphics[width=1.8\columnwidth]{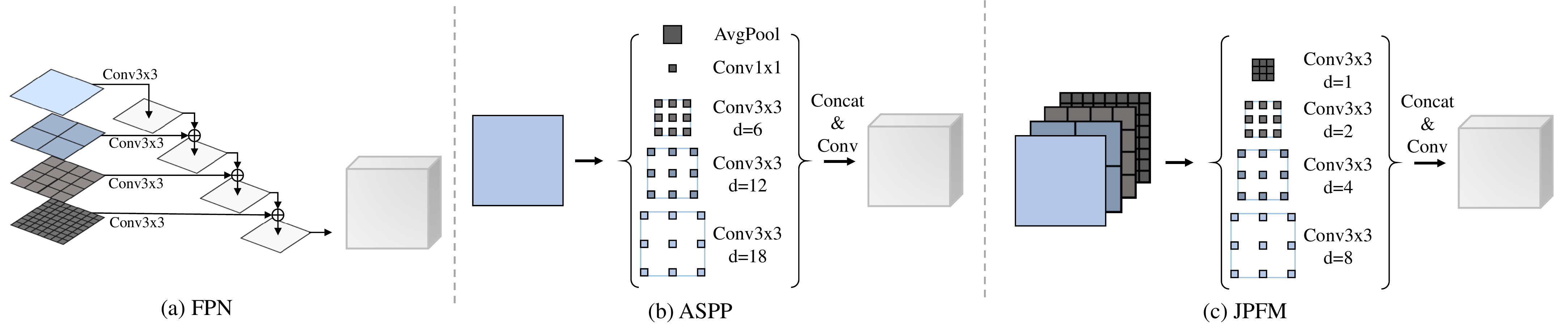}
\caption{\textbf{Comparison among Feature Pyramid Network(FPN), Atrous Spatial Pyramid Pooling (ASPP), and the proposed Joint Pyramid Fusion Module (JPFM).} (a) FPN combines features of different scales by building a feature pyramid and using lateral connections to fuse features from different levels of the pyramid. (b) ASPP utilizes sparse designed dilations~(d = 1, 6, 12, 18) to capture the long-range context from the last layer (Conv5) features.
(c) Multi-scale features from Conv2, Conv3, Conv4, and Conv5 are first upsampled with factors 1, 2, 4, 8, respectively.
Then, JPFM engages dense-designed dilations~(d = 1, 2, 4, 8) to model the cross-scale dependencies, and enhances the task-specific feature for each branch.
}
\label{fig:jpfm}
\end{figure*}

\section{Proposed Method}
The main idea of PointNu-Net is to rephrase the nuclei instance segmentation as two simultaneous prediction problems. Concretely, our system represents one nuclei by a single point at their instance segmentation mask center. The corresponding instance segmentation is then regressed directly using dynamic convolution at the center location. As shown in \autoref{fig:dy}, our method first outputs three predictions, \ie keypoint heatmap, kernel prediction, and feature prediction. The local peaks are then filtered from the keypoint heatmap to locate center points of nuclei. Next, the kernel vectors are selected from the kernel prediction according to the center point position. At last, the instance masks of all the identified center points are produced via dynamically convoluting the corresponding kernel vectors onto the feature prediction.

Our network architecture consists of image feature extractor, feature fusion module, and three task-specific prediction branches, as shown in \autoref{fig:arc}. In the following, we will first describe the prediction branches including the keypoint heatmap regression for nuclei detection and classification as well as the dynamic convolution for nuclei segmentation. Afterward, details of the employed CNN for feature extraction and fusion will be discussed. Finally, we introduce the inference process of our proposed method.

\subsection{Detection And Classification By Keypoints}
Given an input image $I$ of width $W$ and height $H$ and its corresponding ground truth heatmap $Y$, our aim is to regress a low-resolution keypoint heatmap prediction $\hat{Y} \in {[0,1]}^{{\frac{W}{R} \times \frac{H}{R} \times C}}$, where $R$ is the output downsampling factor and $C$ is the number of keypoint types. The default output downsampling factor of $R=4$ is applied following the literature~\cite{stride1,stride2}. Keypoint types include $C=1$ for nuclei detection, or $C>1$ for nuclei detection and classification. A prediction $\hat{Y}_{xyc} = 1$ corresponds to a detected keypoint, whilst $\hat{Y}_{xyc} = 0$ is background, where $x, y$ denote the coordinate and $c$ represents the $c$-th class.

We train the keypoint heatmap prediction branch by following ~\cite{CenterNet}, as shown in \autoref{fig:arc}. We render all ground truth keypoints in a heatmap $\hat{Y} \in {[0,1]}^{{\frac{W}{R} \times \frac{H}{R} \times C}}$ using an unnormalized Gaussian kernel, $\exp{(-\frac{(x-\tilde{p}_x)^2+(y-\tilde{p}_y)^2}{2\sigma^2_p})}$, where $\sigma_p$ is an object size-adaptive standard deviation~\cite{Cornernet}. 
If two Gaussian of the same class overlap, we take the element-wise maximum~\cite{stride1}. The training objective is a penalty-reduced pixel-wise logistic regression with the focal loss~\cite{focalloss}:

\begin{align*}
\mathcal{L}_{keypoint}=\frac{-1}{N_{pos}^k}\sum_{xyc}\left\{\begin{matrix}
 (1-\hat{Y}_{xyc})^\alpha \log{(\hat{Y}_{xyc})}  & \text{if $Y_{xyc}=1$} \\ 
\begin{matrix}
(1-Y_{xyc})^\beta(\hat{Y}_{xyc})^\alpha \\ 
\log{(1-\hat{Y}_{xyc})}
\end{matrix}& \text{otherwise}
\end{matrix}\right.
\end{align*}

where $\alpha$ and $\beta$ are the hyper-parameters of the focal loss and $N_{pos}^k$ is the number of keypoints in image $I$. Normalization by $N_{pos}^k$ is chosen  to normalize all positive focal loss instances to 1. We use $\alpha=2$ and $\beta=4$ in our experiments, following \cite{Cornernet} and~\cite{CenterNet}.

\begin{figure*}[t]
\centering
\includegraphics[width=1.8\columnwidth]{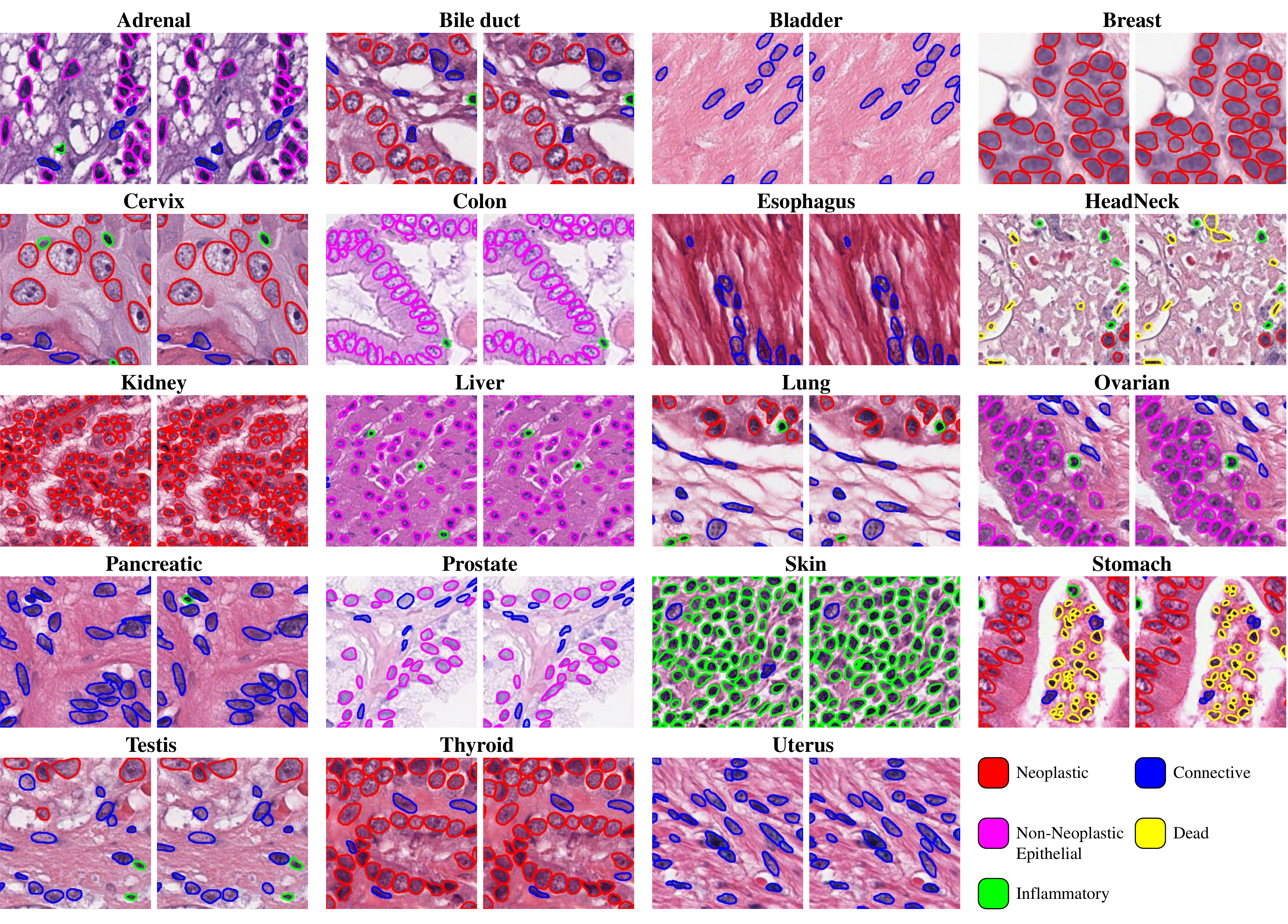}
\caption{\textbf{Examples of PointNu-Net segmentation and classification results across 19 tissues on PanNuke}. In each pair, left is the ground truth overlaid, while the right is the PointNu-Net prediction overlaid. A different color represents a different class.}
\label{fig:pan}
\end{figure*}

\subsection{Instance Segmentation By Dynamic Convolution}
Given the ground truth instance mask set $M$, our goal is to predict the full-size instance mask $\hat{M}_{x,y}$ with size of $W \times H$ for each point at $(x,y)$ on the keypoint heatmap. That is to say, if the center point of one instance mask locates at $(x,y)$ on the keypoint heatmap, this point is responsible to predict that mask. 
Ideally, a one-to-one correspondence can be established between the semantic keypoint heatmap and the class-agnostic mask. However, it may result in unnecessary computing redundancy and out-of-memory problem to directly predict such a tensor. To overcome this problem, we take advantage of dynamic convolution~\cite{condinst,solov2} and divide the mask branch into a feature branch and a kernel branch. For each point in the heatmap, we engage dynamic convolution to produce the final instance mask prediction, which can be written as:
$\hat{M}_{x,y}=K_{x,y}\otimes F$,
where $F \in {\mathbb{R}}^{W \times H \times E}$ is the predicted feature, $K_{x,y} \in {\mathbb{R}}^{1 \times 1 \times E}$ is the convolution kernel generated by kernel branch, $\mathbb{R}$ is the field of real numbers, $E$ is the length of kernel size, and $\otimes$ is the convolution operation. 

With the decoupling of the feature and kernel predictions, it is no longer necessary to predict the mask for all points. Instead, only the kernels corresponding to the positives are selected to compute the final prediction and calculate the training loss. Inspired by Wang~\etal\cite{solo}, we add locations in the neighborhood of the center as positives where the corresponding heatmap is larger than a threshold $\tau$ (\eg $ \rm{heatmap}>\tau$). The training objective of the feature branch and kernel branch is a Soft Dice loss~\cite{vnet}
\begin{equation*}
\mathcal{L}_{mask}=\frac{1}{N_{pos}^m}\sum_{xyk}\mathbbm{1}_{(\underset{c \in C}{\max}(Y_{xyc}^k)>\tau)}(1-\rm{DICE}( \hat{M}_{x,y},M^k)),
\end{equation*}
where $N_{pos}^m$ is the number of instance masks, $\rm DICE$ is Dice coefficient, $\hat{M}_{x,y}$ is the predicted soft mask, $k$ is the $\rm k^{\text{th}}$ instance in image $I$, and ${M}^{k}$ is the ground truth mask corresponding to $\rm k^{\text{th}}$ instance. $\mathbbm{1}$ is the indicator function, being 1 if $\underset{c \in C}{\max}(Y_{xyc})>\tau$ and 0 otherwise.

The overall objective is:
\begin{equation*}
\mathcal{L}_{total}=\mathcal{L}_{keypoint}+\lambda_{mask}\mathcal{L}_{mask}.
\end{equation*}
We set $\lambda_{mask}=1$ and $\tau=0.5$ empirically in all our experiments unless specified otherwise. We apply a single network to predict semantic keypoint $\hat{Y}$, dynamic convolution kernel $K$ and feature $F$ simultaneously with three separated branches (shown in \autoref{fig:arc}), and generate class-agnostic instance masks $\hat{M}$ dynamically. 

\subsection{Feature Extraction And Fusion}
Standard deep-learning-based feature extraction, like ResNet~\cite{he2016identity} and VGGNet~\cite{simonyan2018very}, cannot extract a strong and representative set of features for small objects, since they encode the input image as a high-level low-resolution representation in a series manner. However, the size of nuclei varies among the tissues, and many nuclei are very small (less than $8\times 8$ pixels). This may result in unsatisfactory detection and segmentation  for small nuclei. To deal with this problem, we take High Resolution Net (HRNet)~\cite{hrnet} as our backbone which can maintain high-resolution representations as well as extract low-resolution high-level features throughout the whole process. As shown in \autoref{fig:arc}, multi-scale feature maps are obtained with HRNet for further prediction (it is noted that HRNet structure is omitted in this figure for clarity, which is detailed in the supplementary). In particular, the output feature maps from stage Conv2 to Conv5 are extracted from HRNet, where Conv$L$ has the resolution $2^L$ lower than the input image. Different scales of features are upsampled to the same size as Conv2 using bilinear interpolation and concatenated together.

\begin{table*}[ht]\small
\setlength\tabcolsep{4.5 pt}
\centering
\caption{Comparative experiments of nuclei segmentation and classification on the PanNuke.}
\begin{tabular}{lcccccccccccclllcc}
\toprule
\multirow{2}{*}{Tissue} & \multicolumn{2}{l}{DIST} & \multicolumn{1}{l}{} & \multicolumn{2}{l}{Mask-RCNN} & \multicolumn{1}{l}{} & \multicolumn{2}{l}{Micro-Net} & \multicolumn{1}{l}{} & \multicolumn{2}{l}{HoVer-Net} & \multicolumn{1}{l}{} & \multicolumn{2}{l}{ CPP-Net } &  & \multicolumn{2}{l}{PointNu-Net} \\ \cmidrule{2-3} \cmidrule{5-6} \cmidrule{8-9} \cmidrule{11-12} \cmidrule{14-15} \cmidrule{17-18} 
 & \multicolumn{1}{l}{mPQ} & \multicolumn{1}{l}{bPQ} & \multicolumn{1}{l}{} & \multicolumn{1}{l}{mPQ} & \multicolumn{1}{l}{bPQ} & \multicolumn{1}{l}{} & \multicolumn{1}{l}{mPQ} & \multicolumn{1}{l}{bPQ} & \multicolumn{1}{l}{} & \multicolumn{1}{l}{mPQ} & \multicolumn{1}{l}{bPQ} & \multicolumn{1}{l}{} & mPQ & bPQ &  & \multicolumn{1}{l}{mPQ} & \multicolumn{1}{l}{bPQ} \\ \midrule
Adrenal Gland & 0.3442 & 0.5603 &  & 0.347 & 0.5546 &  & 0.4153 & 0.644 &  & 0.4812 & 0.6962 &  & 0.4922 & 0.7031 &  & \textbf{0.5115} & \textbf{0.7134} \\
Bile Duct & 0.3614 & 0.5384 &  & 0.3536 & 0.5567 &  & 0.4124 & 0.6232 &  & 0.4714 & 0.6696 &  & 0.4650 & 0.6739 &  & \textbf{0.4868} & \textbf{0.6814} \\
Bladder & 0.4463 & 0.5625 &  & 0.5065 & 0.6049 &  & 0.5357 & 0.6488 &  & 0.5792 & 0.7031 &  & 0.5932 & 0.7057 &  & \textbf{0.6065} & \textbf{0.7226} \\
Breast & 0.3790 & 0.5466 &  & 0.3882 & 0.5574 &  & 0.4407 & 0.6029 &  & 0.4902 & 0.6470 &  & 0.5066 & \textbf{0.6718} &  & \textbf{0.5147} & 0.6709 \\
Cervix & 0.3371 & 0.5309 &  & 0.3402 & 0.5483 &  & 0.3795 & 0.6101 &  & 0.4438 & 0.6652 &  & 0.4779 & 0.6880 &  & \textbf{0.5014} & \textbf{0.6899} \\
Colon & 0.2989 & 0.4508 &  & 0.3122 & 0.4603 &  & 0.3414 & 0.4972 &  & 0.4095 & 0.5575 &  & 0.4269 & 0.5888 &  & \textbf{0.4509} & \textbf{0.5945} \\
Esophagus & 0.3942 & 0.5295 &  & 0.4311 & 0.5691 &  & 0.4668 & 0.6011 &  & 0.5085 & 0.6427 &  & 0.5410 & 0.6755 &  & \textbf{0.5504} & \textbf{0.6766} \\
Head \& Neck & 0.3177 & 0.4764 &  & 0.3946 & 0.5457 &  & 0.3668 & 0.5242 &  & 0.4530 & 0.6331 &  & 0.4667 & 0.6468 &  & \textbf{0.4838} & \textbf{0.6546} \\
Kidney & 0.3339 & 0.5727 &  & 0.3553 & 0.5092 &  & 0.4165 & 0.6321 &  & 0.4424 & 0.6836 &  & \textbf{0.5092} & \textbf{0.7001} &  & 0.5066 & 0.6912 \\
Liver & 0.3441 & 0.5818 &  & 0.4103 & 0.6085 &  & 0.4365 & 0.6666 &  & 0.4974 & 0.7248 &  & 0.5099 & 0.7271 &  & \textbf{0.5174} & \textbf{0.7314} \\
Lung & 0.2809 & 0.4978 &  & 0.3182 & 0.5134 &  & 0.337 & 0.5588 &  & 0.4004 & 0.6302 &  & \textbf{0.4234} & \textbf{0.6364} &  & 0.4048 & 0.6352 \\
Ovarian & 0.3789 & 0.5289 &  & 0.4337 & 0.5784 &  & 0.4387 & 0.6013 &  & 0.4863 & 0.6309 &  & 0.5276 & 0.6792 &  & \textbf{0.5484} & \textbf{0.6863} \\
Pancreatic & 0.3395 & 0.5343 &  & 0.3624 & 0.5460 &  & 0.4041 & 0.6074 &  & 0.4600 & 0.6491 &  & 0.4680 & 0.6742 &  & \textbf{0.4804} & \textbf{0.6791} \\
Prostate & 0.3810 & 0.5442 &  & 0.3959 & 0.5789 &  & 0.4341 & 0.6049 &  & 0.5101 & 0.6615 &  & \textbf{0.5261} & \textbf{0.6903} &  & 0.5127 & 0.6854 \\
Skin & 0.2627 & 0.5080 &  & 0.2665 & 0.5021 &  & 0.3223 & 0.5817 &  & 0.3429 & 0.6234 &  & 0.3547 & 0.6192 &  & \textbf{0.4011} & \textbf{0.6494} \\
Stomach & 0.3369 & 0.5553 &  & 0.3684 & 0.5976 &  & 0.3872 & 0.6293 &  & \textbf{0.4726} & 0.6886 &  & 0.4553 & \textbf{0.7043} &  & 0.4517 & 0.7010 \\
Testis & 0.3278 & 0.5548 &  & 0.3512 & 0.5420 &  & 0.4088 & 0.6300 &  & 0.4754 & 0.6890 &  & 0.4917 & 0.7006 &  & \textbf{0.5334} & \textbf{0.7058} \\
Thyroid & 0.2574 & 0.5596 &  & 0.3037 & 0.5712 &  & 0.3712 & 0.6555 &  & 0.4315 & 0.6983 &  & 0.4344 & \textbf{0.7094} &  & \textbf{0.4508} & 0.7076 \\
Uterus & 0.3487 & 0.5246 &  & 0.3683 & 0.5589 &  & 0.3965 & 0.5821 &  & 0.4393 & 0.6393 &  & 0.4790 & 0.6622 &  & \textbf{0.4846} & \textbf{0.6634} \\ \hline
Average  & 0.3406 & 0.5346 &  & 0.3688 & 0.5528 &  & 0.4059 & 0.6053 &  & 0.4629 & 0.6596 &  & 0.4817 & 0.6767 &  & \textbf{0.4957} & \textbf{0.6808}\\\bottomrule
\end{tabular}
\label{tab:pannuke}
\end{table*}

\begin{table*}[t]
\setlength\tabcolsep{4 pt}
\centering
\caption{Precision (P), Recall (R) and F1-score (F1) across three dataset splits for detection and classification of nuclei types on PanNuKe.}
\begin{tabular}{lllllllllllllllllllllllll}
\toprule
\multicolumn{4}{c}{\multirow{2}{*}{Detection}} &  & \multicolumn{19}{c}{Classification} &  \\ \cmidrule{6-24}
\multicolumn{4}{c}{} &  & \multicolumn{3}{l}{Neoplastic} &  & \multicolumn{3}{l}{Non-Neoplastic} &  & \multicolumn{3}{l}{Inflammatory} &  & \multicolumn{3}{l}{Connective} &  & \multicolumn{3}{l}{Dead} &  \\ \cmidrule{6-8} \cmidrule{10-12} \cmidrule{14-16} \cmidrule{18-20} \cmidrule{22-24}
 & P & R & F1 &  & P & R & F1 &  & P & R & F1 &  & P & R & F1 &  & P & R & F1 &  & P & R & F1 &  \\ \midrule
DIST & 0.74 & 0.71 & 0.73 &  & 0.49 & 0.55 & 0.50 &  & 0.38 & 0.33 & 0.35 &  & 0.42 & 0.45 & 0.42 &  & 0.42 & 0.37 & 0.39 &  & 0.00 & 0.00 & 0.00 &  \\
M-RCNN & 0.76 & 0.68 & 0.72 &  & 0.55 & 0.63 & 0.59 &  & 0.52 & 0.52 & 0.52 &  & 0.46 & 0.54 & 0.50 &  & 0.42 & 0.43 & 0.42 &  & 0.17 & 0.30 & 0.22 &  \\
Micro-Net & 0.78 & \textbf{0.82} & 0.80 &  & 0.59 & 0.66 & 0.62 &  & 0.63 & 0.54 & 0.58 &  & 0.59 & 0.46 & 0.52 &  & 0.50 & 0.45 & 0.47 &  & 0.23 & 0.17 & 0.19 &  \\
Hover-Net & \textbf{0.82} & 0.79 & 0.80 &  & 0.58 & 0.67 & 0.62 &  & 0.54 & 0.60 & 0.45 &  & 0.56 & 0.51 & 0.54 &  & 0.52 & 0.47 & 0.49 &  & 0.28 & \textbf{0.35} & \textbf{0.31} &  \\
PointNu-Net & 0.81 & 0.81 & \textbf{0.81} &  & \textbf{0.74} & \textbf{0.72} & \textbf{0.73} &  & \textbf{0.75} & \textbf{0.71} & \textbf{0.73} &  & \textbf{0.63} & \textbf{0.57} & \textbf{0.60} &  & \textbf{0.61} & \textbf{0.55} & \textbf{0.58} &  & \textbf{0.45} & 0.24 & \textbf{0.31} & \\\bottomrule
\end{tabular}
\label{tab:detcls}
\end{table*}

\begin{table}[ht]\small
\setlength\tabcolsep{4.5 pt}
\centering
\caption{Average PQ across three dataset splits on PanNuke for each nuclear category}
\begin{tabular}{lccccc}
\toprule
 & Neo & Non-Neo Epi & Inflam & Conn & Dead \\ \midrule
DIST & 0.439 & 0.290 & 0.343 & 0.275 & 0.000 \\
Mask-RCNN & 0.472 & 0.403 & 0.290 & 0.300 & 0.069 \\
Micro-Net & 0.504 & 0.442 & 0.333 & 0.334 & 0.051 \\
HoVer-Net & 0.551 & 0.491 & 0.417 & 0.388 & 0.139 \\
PointNu-Net & \textbf{0.578} & \textbf{0.577} & \textbf{0.433} & \textbf{0.409} & \textbf{0.154}\\\bottomrule
\end{tabular}
\label{tab:nt}
\end{table}

Rather than using the most commonly-used feature fusion FPN to better integrate the local features from the backbone for nuclei detection and classification, we design a simple but efficient module called JPFM to make sure each branch can make full use of the maintained high-resolution representations from the backbone. Inspired by ASPP~\cite{chen2017deeplab}, we engage a dense design of dilated convolution ($d=1, 2, 4, 8$) to extract task-specific representation for each branch. The difference is that, ASPP utilized sparse dilation (e.g., $d=1, 6, 12, 18$) rates to capture the long-range context from the last layer feature, while our proposed JPFM employs dense dilation rates aiming to model cross-scale dependencies among the features from Conv2 to Conv5 and enhance the task-related features (see \autoref{fig:jpfm} for a detailed comparison). With this novel design, JPFM proves able to  better exploit the maintained high-resolution representations, leading to much higher performance compared with ASPP, as later shown in the experiments in Sec.~\ref{sec:neck}.

Finally, \textit{Conv Blocks} are stacked in each branch to generate the final prediction, while \textit{Upsample Blocks} are used in the feature branch to output high-resolution features. Followed by \cite{solov2}, normalized coordinates are added at the first \textit{Conv Blocks} of the kernel branch to embed position information.

\subsection{From Points To Instance Segmentation}
At inference time, as shown in \autoref{fig:dy}, we first extract the peaks in the heatmap for each category independently. We detect all responses whose value is greater or equal to its 8-connected neighbors and a confidence threshold of 0.4 is used to filter out predictions with low confidence. Then the local peaks and their corresponding kernels are selected, and dynamic convolution is performed between the predicted feature and selected kernels to obtain the binary masks. Matrix NMS operation~\cite{solov2} is alternatively used to get rid of overlapped predictions, which can slightly improve the overall results.

\section{Experiments}
\subsection{Datasets}
\noindent\textbf{PanNuke} is so far the largest publicly available nuclei segmentation and classification dataset~\cite{gamper2019pannuke,gamper2020pannuke}.
Images with 19 different tissues were obtained from The Cancer Genome Atlas (TCGA), where $256\times 256$ pre-extract patches were collected from more than 20,000 Whole Slide Images. 
PanNuke is split into 3 randomized training, validation, and testing folds, and we follow~\cite{gamper2020pannuke} to reproduce the results. This dataset is semi-automatically annotated and quality controlled by clinical pathologists, which is statistically similar to ‘the clinical wild’ and with minimal selection bias.\\
\textbf{Kumar} is a common nuclei segmentation dataset~\cite{kumar} consisting of 30 images from seven different tissues. They are divided into a training set of 16 images (4 breast, 4 liver, 4 kidney and 4 prostate) and a testing set of 14 images (2 breast, 2 liver, 2 kidney and 2 prostate, 2 bladder, 2 colon, 2 stomach) with the same protocol used in~\cite{hovernet,kumar,dist}.\\
\textbf{CoNSeP}~\cite{hovernet} containing 41 images with different cell types. Four classes are considered: epithelial, inflammatory, spindle-shaped, and miscellaneous. The training and test set partition follows  previous works, where the training set contains 27 images, and the test set contains 14 images.

\begin{figure}[t] 
\centering
\includegraphics[width=0.75\columnwidth]{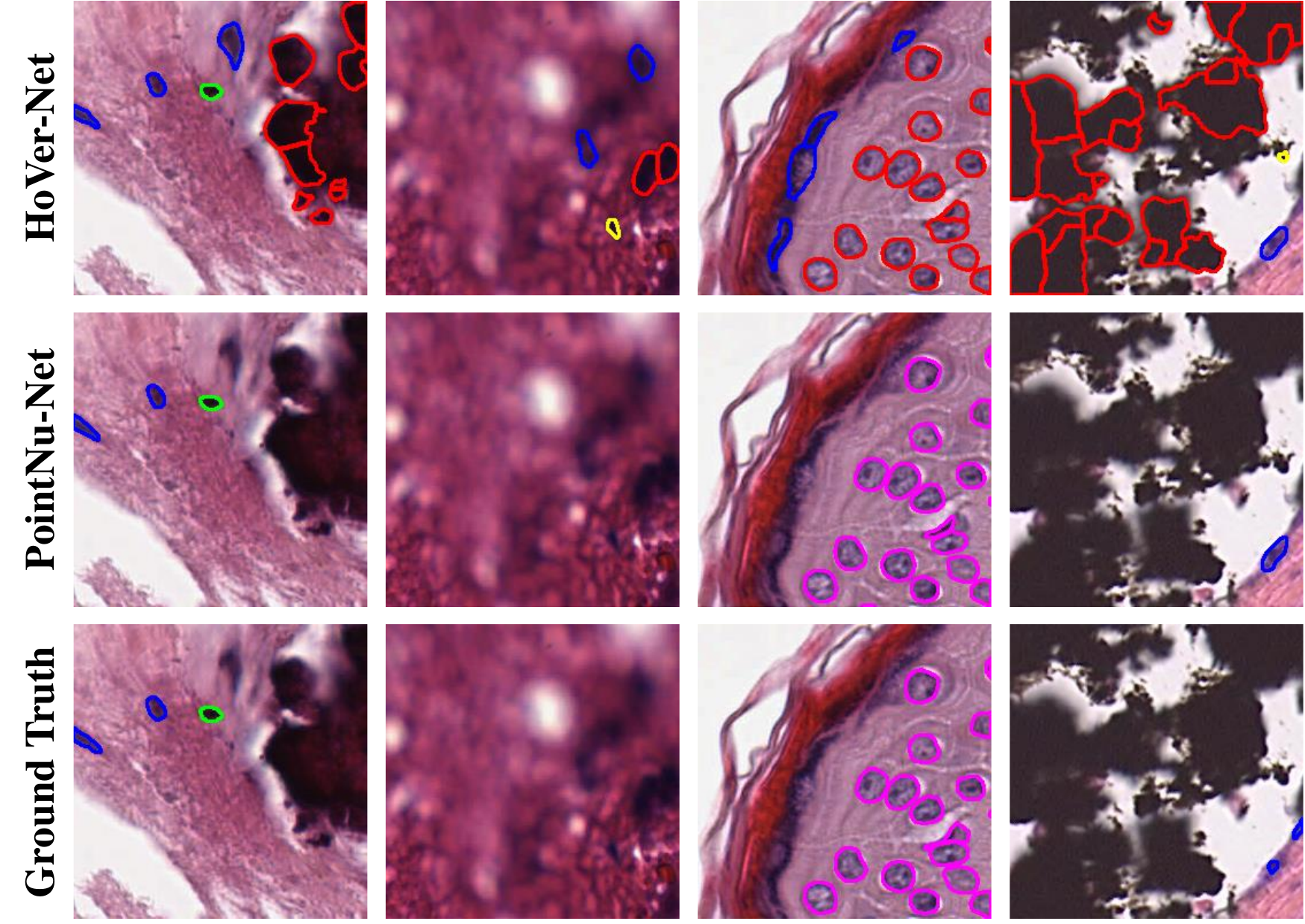}
\caption{\textbf{A selection of visual fields from PanNuke which are complicated scenarios in 'the clinical wild'.}}
\label{fig:cwild}
\end{figure}

\begin{table*}[ht]\footnotesize
\centering
\caption{Comparative experiments on small datasets Kumar and CoNSeP in terms of detection (Det), segmentation (Seg), and classification (Cls). bPQ reflects the performance on nuclei detection and segmentation, while mPQ highlights the performance on nuclei detection, segmentation and classification. $\rm{F_d}$ denotes the F1-score for nuclear detection, whereas $\rm{F_c^e}$, $\rm{F_c^i}$, $\rm{F_c^s}$ and $\rm{F_c^m}$ denote the F1 classification score for the epithelial, inflammatory, spindle-shaped and miscellaneous classes respectively.}
\begin{tabular}{lllllllllllllllll}
\toprule
   & \multicolumn{3}{l}{Kumar(Det+Seg)} &  & \multicolumn{3}{l}{CoNSeP (Det+Seg)} &  & \multicolumn{6}{l}{CoNSeP (Det+Seg+Cls)}    \\ \cmidrule{2-4} \cmidrule{6-8} \cmidrule{10-15}
   & DQ & SQ & bPQ &  &  DQ & SQ & bPQ && mPQ & $\rm{F_d}$ & $\rm{F_c^e}$ & $\rm{F_c^i}$ & $\rm{F_c^s}$ & \rm{$F_c^m$}  \\ \midrule
U-Net~\cite{unet} &0.691&0.690&0.478&&0.488&0.671&0.328& & -  &-&-&-&-&-\\
DIST~\cite{dist}  & 0.601 & 0.732 & 0.443 &  &  0.544 & 0.728 & 0.398 && 0.372  & 0.712 & 0.617 & 0.534 & 0.505 & 0.000\\
Mask-RCNN~\cite{maskrcnn}  & 0.704 & 0.720 & 0.509 &  &  0.619 & 0.740 & 0.460 &&0.450   & 0.692 & 0.595 & 0.590 & 0.520 & 0.098\\
Micro-Net~\cite{micronet}   & 0.692 & 0.747 & 0.519 &   & 0.600 & 0.745 & 0.449 && 0.430 & 0.743 & 0.615 & 0.592 & 0.532 & 0.117 \\
CIA-Net~\cite{cianet}   & 0.754 & 0.762 & 0.577 &   & - & - & -& &-   &-&-&-&-&-\\
HoVer-Net~\cite{hovernet}   & 0.770 & \textbf{0.773} & 0.597 &   & 0.702 & \textbf{0.778} & 0.547 && 0.516 & 0.748 & 0.635 & 0.631 & \textbf{0.566} & 0.426 \\
Triple U-Net~\cite{tripleunet}	 & - & - & 0.601 &  & - & - & \textbf{0.562} &&- &-&-&-&-&-\\
PointNu-Net   & \textbf{0.784} & 0.768 & \textbf{0.603} &  & \textbf{0.714} & 0.762 & 0.555 && \textbf{0.536} & \textbf{0.752} & \textbf{0.661} & \textbf{0.647} & 0.559 & \textbf{0.462} \\ \bottomrule
\end{tabular}
\label{tab:kumar}
\end{table*}

\begin{figure*}[ht] 
\centering
\includegraphics[width=1.8\columnwidth]{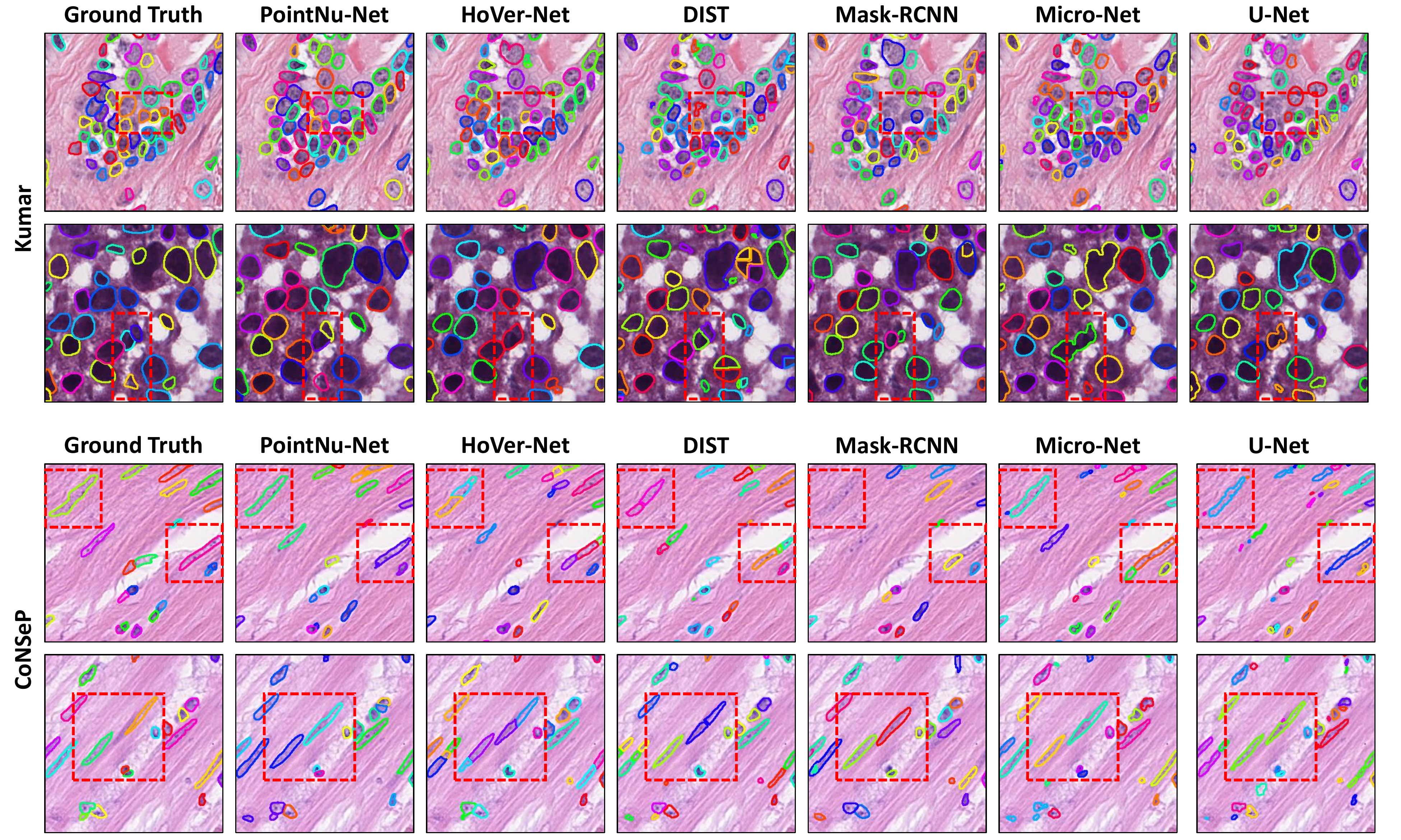}
\caption{Examples of visual nuclei segmentation results on Kumar and CoNSeP. For each dataset, we displayed the 6 models from left to right. The red dashed boxes indicate the better detection performance of PointNe-Net. The different colors of the nuclear boundaries denote separate instances. }
\label{fig:visual1}
\end{figure*}

\subsection{Implementation And Training Details}
We implemented our framework with the open-source software library PyTorch 1.8.0 on a workstation equipped with one NVIDIA GeForce RTX 3090 GPU. For Kumar dataset, we used stain normalization~\cite{stainnorm} to reduce the color differences between the stained images, and no stain normalization was performed for CoNSeP and PanNuke since the color difference between training and testing data is not large. The training objective function consists of the keypoint regression loss $\mathcal{L}_{keypoint}$ and instance segmentation loss $\mathcal{L}_{mask}$, and the weights for each loss are both set to 1. AdamW~\cite{adamw} has been used as an optimizer to minimize objective function with the mini-batch of 8 and weight decay of 0.0001. All models are trained for 100 epochs with an initial learning rate of 0.0001, which is then divided by 10 at the $80^{\text{th}}$ and again at the $90^{\text{th}}$ epoch. Various data augmentation techniques were employed including random cropping, flipping, color jittering, blurring and elastic transformation by following HoVer-Net~\cite{hovernet}. During the training phase, image patches with the resolution of 256 $\times$ 256 are randomly cropped, while during inference, we adopt the sliding window strategy used in previous works~\cite{dist,cianet,hovernet} to handle images of any size.

\subsection{Results And Comparative Analysis}
To quantify the instance segmentation performance of each method, we used panoptic quality (PQ), followed by Graham~\etal\cite{hovernet}. Panoptic quality was further tear apart into Detection Quality (DQ) and Segmentation Quality (SQ) components for interpretability, which is defined as follows:
\begin{equation*}
\small
{\rm PQ}= \underbrace{\frac{|TP|}{ |TP|+\frac{1}{2}|FP| + \frac{1}{2}|FN|}}_{\text{Detection quality (DQ)}} \times \underbrace{\frac{\sum_{(p,g)\in TP} {\rm IOU}(p,g)}{ |TP|}}_{\text{Segmentation quality (SQ)}}.
\end{equation*}
We used multi-class PQ (\textbf{mPQ}) to evaluate the performance of detection, segmentation and classification, while binary PQ (\textbf{bPQ}) was used to evaluate the performance of detection and segmentation. Specifically, mPQ was calculated independently for each positive class, and bPQ assumes that all nuclei belong to one class.

On large dataset PanNuke,  the proposed method was compared against several deep-learning-based methods, \eg DIST~\cite{dist}, Mask-RCNN~\cite{maskrcnn}, Micro-Net~\cite{micronet}, HoVer-Net\cite{hovernet} and CPP-Net~\cite{cppnet}. As shown in \autoref{tab:pannuke}, the proposed method achieves state-of-the-art performance not only on classification but also on segmentation. In particular, PointNu-Net outperformed the third-best method HoVer-Net and second-best method CPP-Net, by 0.0328 and 0.0140 on the average mPQ across tissues respectively. To take an insight into how PointNu-Net performed for different types of nuclei, we reported mPQ and bPQ for all 19 tissue types separately. We observed that for most tissue types, PointNu-Net achieved the best performance against all the existing methods. We also reported the average PQ for each type of nuclei in \autoref{tab:nt}. In some cases, distinguishing between neoplastic and non-neoplastic nuclei proved to be challenging, yet an improvement of 0.027 and 0.086 on the average PQ on both nuclei can be observed in our method. 
Meanwhile, \autoref{tab:detcls} delineates the detection and classification performance for different nuclei types on PanNuKe. As observed, in detection, our PointNu-Net just showed marginal improvement over HoverNet (0.81 \textit{vs.} 0.80 in terms of F1); however, in classification, our proposed model attained considerable enhancements, specifically for Neoplastic and Non-Neoplastic cells. This validated the superior  ability of our model in classification.
In addition, when facing complicated scenarios, e.g., out-of-focus, blur boundaries, and imbalanced staining, previous methods, e.g., HoVer-Net, might perform poorly since there is no confidence mechanism to filter the poor instance results. In contrast, our method outputted instances with high confidence, resulting in a more robust performance in 'the clinical wild', as shown in \autoref{fig:cwild}.

On small datasets Kumar and CoNSeP, we evaluated the performance of instance segmentation with the above-mentioned methods. We also made comparisons with some additional methods, including U-Net~\cite{unet}, CIA-Net~\cite{cianet}, and Triple U-Net~\cite{tripleunet}.  As observed from the results in \autoref{tab:kumar}, our method generated the highest detection quality (DQ) on Kumar and CoNSeP, owing to PointNu-Net's strong detection capability. On the other hand, though PointNu-Net still performs competitively, it is observed that our method resulted in marginally lower segmentation quality (SQ) than HoVer-Net. This can be explained by the fact that our method only outputs the instances with high confidence whilst those with low confidence may be ignored. Finally, in terms of the overall quality bPQ, the proposed PointNu-Net performs the best on Kumar dataset while it is slightly worse than  HoVer-Net on CoNSep.

In addition, we also examined nuclei detection, segmentation, and classification. Since no classification annotation is conducted in Kumar, we only report the result on  CoNSeP. As shown in the right part of \autoref{tab:kumar}, the proposed method leads to the  superior performance. The post-processing may somehow limit the bottom-up methods' performance since additive semantic segmentation needs to be handled  for classification results. In comparison, our keypoint-aware framework brings an overall improvement of 0.02 mPQ on CoNSeP over the best result given by HoVer-Net. It is worth noting that the detection only yielded a marginal improvement of 0.004 in terms of $F_d$ on the CoNSeP dataset. That is to say, the main improvement of our framework is on the classification in this dataset, which has an average increase of 0.018 on F1 classification scores for four classes.  

Some qualitative results are visualized for large datasets in \autoref{fig:pan}. The proposed PointNu-Net has shown impressive results on nuclei detection, segmentation and classification across 19 tissue on PanNuKe. Meanwhile, PointNu-Net can identify individual nuclei in some hard cases. As shown in the red dashed boxes at the third and fourth row of \autoref{fig:visual1}, the bottom-up methods, e.g., HoVer-Net and DIST, tend to identify one nuclei with strip shape as more than one nuclei, since the post-processing for those methods are designed for regular nuclei, and the manually defined separating rules for nuclei cannot cover all the circumstances. By contrast, our method can detect such nuclei well since no post-processing is required to distinguish instances.

\subsection{Speed-accuracy Trade-off}
Although it is rarely mentioned in the previous works, the speed of processing images  also matters due to the extra large size of images in WSIs. In order to investigate the trade-off between the efficiency and the accuracy quantitatively, we performed segmentation and classification on datasets CoNSeP over 10 runs and calculated the average inference time on a $1,000 \times 1,000$ image, as shown in \autoref{tab:tf}. As a region proposal-based method that will generate redundant and overlapped region-of-interests, the top-down method Mask-RCNN required larger space to store intermediate results and a longer time to post-process, resulting in almost 52 seconds of processing time per image. The bottom-up method HoVer-Net took about 5.3 seconds per image thanks to its parallel separation of instances. The post-processing of HoVer-Net contained complex operations using CPU, which is a bottleneck for faster inference time. PointNu-Net detected each instance using heatmap peaks and segmented instances dynamically without any post-processing, which takes advantage of parallel calculation on GPU. As a consequence, our default version of PointNu-Net spent 3.29 seconds of inference time to achieve the highest mPQ on CoNSeP, approximately 1.6 $\times$ faster than HoVer-Net. On the other hand, the inference time declined about 10\% without non-maximum suppression, just leading to a slight decrease in performance. Meanwhile, because almost 95\% inference time was costed on GPU, using a smaller model can significantly speed up the model. Here we propose PointNu-Net-M and PointNu-Net-S, which used HRNet-w32 and HRNet-w18 as the backbone, respectively. Both  lighter versions of PointNu-Net share the same training protocol and architectures as the default version, except using 4 instead of 7 stacked convolutional layers on the keypoint heatmap branch and kernel branch. Eventually, we can halve the inference time compared with the default PointNu-Net, while achieving acceptable performance.

\begin{table}[]\footnotesize
\centering
\caption{Speed-accuracy trade-off on the dataset CoNSeP.  $^\dagger$ denotes model inference without matrix NMS operation.}
\begin{tabular}{lcccl}
\toprule
 & Params  & Running Time & \multicolumn{1}{c}{mPQ} \\ \midrule
Mask-RCNN & 43.9 M & 51.97 s & 0.450 \\
HoVer-Net & 54.7 M & 5.29 s & 0.516 \\
PointNu-Net & 160.4 M & 3.29 s & \textbf{0.536} \\
PointNu-Net$^\dagger$ & 160.4 M & 3.07 s & 0.530 \\
PointNu-Net-M & 57.2 M & 2.50 s & 0.530 \\
PointNu-Net-M$^\dagger$  & 57.2 M & 2.24 s & 0.527 \\
PointNu-Net-S  & 26.0 M & \textbf{1.71 s} & 0.518 \\ \bottomrule
\end{tabular}
\label{tab:tf}
\end{table}

\subsection{Ablation Study}
We conducted ablation studies to gain a full understanding of the main components in PointNu-Net. In order to get convincing results, we evaluate all the experiments on the large dataset PanNuke.
 
 \begin{figure}[t]
\centering
\includegraphics[width=0.75\columnwidth]{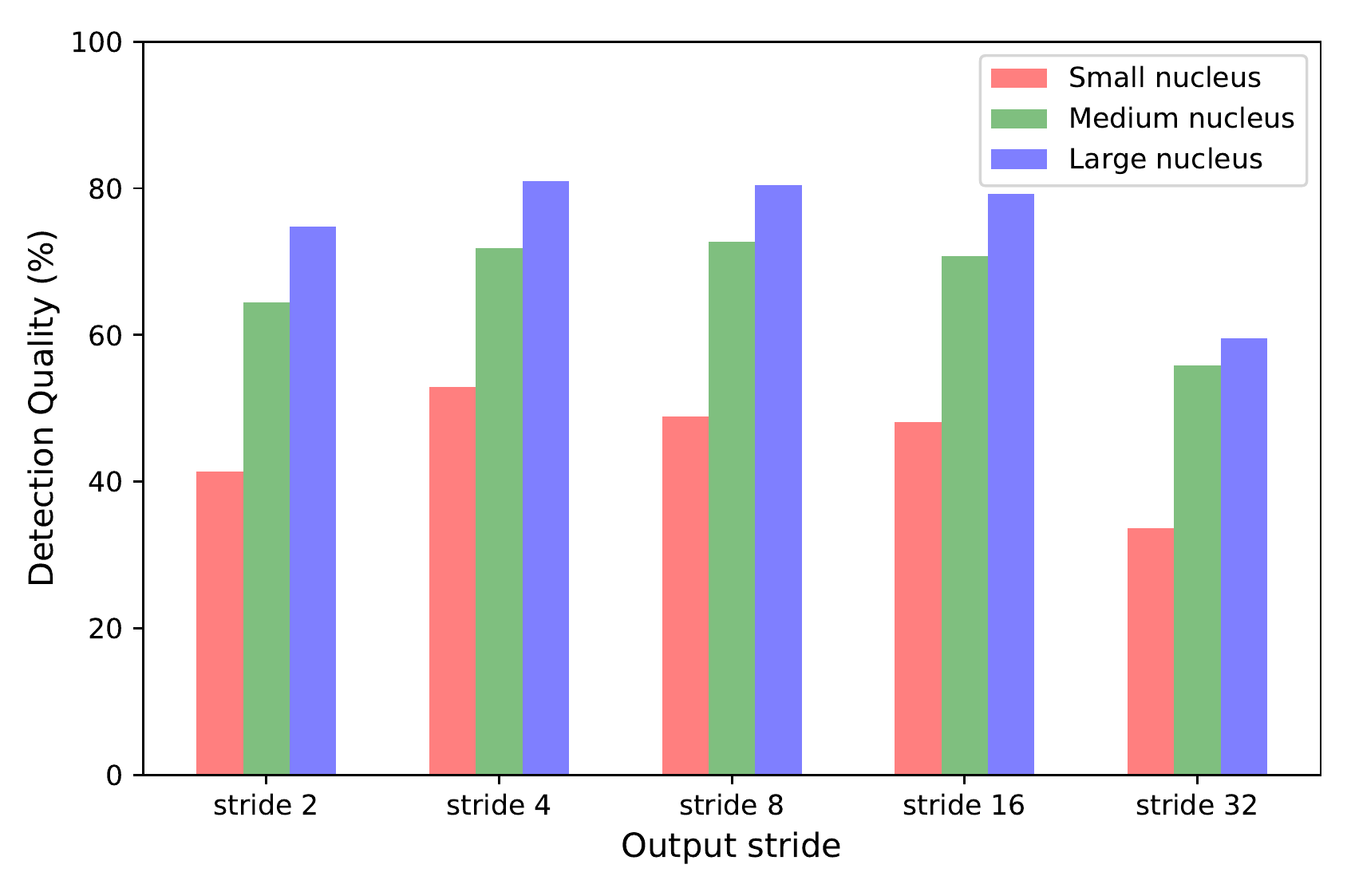}
\caption{Ablation experiments on output stride for keypoint heatmap branch and kernel branch. The bar graph demonstrates the detection performance of PointNu-Net for small, medium, and large nuclei under different output strides.}
\label{fig:stride}
\end{figure}

\subsubsection{Output Downsampling Factor}
We conduct several experiments to investigate how the output downsampling factor $R$ in the keypoint heatmap branch and kernel branch affects the detection performance of PointNu-Net, especially for small nuclei. As shown in \autoref{fig:stride}, after ranking the sizes of the nuclei in PanNuKe dataset from small to large, we group the first 25\% and last 25\% nuclei as small  and large nuclei respectively and leave the rest as medium. \autoref{fig:stride} presents the detection performance comparison under output downsampling factor $R$ of \{2, 4, 8, 16, 32\}. In general, it can be noticed that the most challenging task is to detect small nuclei. With the increasing size of nuclei, nuclei tend to be detected more easily. When $R = 2$, a maximum detection resolution can be reached. Contrary to expectations, omnidirectional performance degradation can be observed, which may be caused by the increasing training difficulty and a worse class imbalance between positive and negative points in the keypoint heatmap. When we use the default setting $R = 4$ for PointNu-Net, the detection of small nuclei reaches the peak. When we further increase $R$, the detection resolution is not high enough to detect all touching small nuclei, thus lowering the detection performance of detecting small nuclei as well as the overall performance. To conclude, the output downsampling factor $R = 4$ is considered better for the keypoint heatmap branch and kernel branch in PointNu-Net.

\subsubsection{Keypoint-aware Detector and Dynamic Segmentor}
\label{sec:abla1}

Here, we investigate the importance of keypoint regression and focal loss for nuclei detection and classification, as well as dynamic segmentation for nuclei segmentation.
As illustrated in \autoref{tab:abkey}, we initiate our analysis by comparing the performance of two loss functions, Binary Cross Entropy (BCE) loss and Focal loss, when applied to both Centerpoint Map and Keypoint Heatmap regression targets. Focal loss consistently outperforms BCE loss in terms of mPQ and bPQ, showing its effectiveness in addressing class imbalance and fostering better training convergence.
Next, we assess the contributions of Keypoint heatmap regression. The Centerpoint Map refers to regression targets containing only the center points of nuclei. Keypoint heatmap regression leads to a significant performance boost in both mPQ (0.032) and bPQ (0.038) compared to Centerpoint Map. This improvement is attributed to its ability to create smoother ground truth around center points, which simplifies the prediction task.
Additionally, we verify the effectiveness of dynamic segmentation by using a single segmentation branch with the standard convolution to generate instance masks for all positions. Dynamic segmentation contributes to better performance by 0.012 (mPQ) and 0.022 (bPQ) by diminishing computational redundancy for negative samples and enhancing overall efficiency.
The synergy of keypoint heatmap regression, focal loss, and dynamic segmentation results in the best overall performance, with an mPQ of 0.496 and a bPQ of 0.681. This demonstrates the effectiveness of these three components within the PointNu-Net framework for nuclei segmentation and classification, highlighting that the three components mutually reinforce one another and are all essential for achieving superior domain adaptation results.

\begin{table}[t]
\setlength\tabcolsep{4.5 pt}
\centering
\caption{Effectiveness of keypoint-aware detector and dynamic segmentation.}
\begin{tabular}{cclccc}
\toprule
\multicolumn{2}{c}{Detection \& Classification} &  & Segmentation & \multirow{2}{*}{mPQ} & \multirow{2}{*}{bPQ} \\ \cmidrule{1-2} \cmidrule{4-4}
Regression Target & Loss Func. &  & Method &  &  \\ \midrule
Centerpoint Map & BCE &  & Standard & 0.401 & 0.565 \\
Centerpoint Map & BCE &  & Dynamic & 0.431 & 0.595 \\
Centerpoint Map & Focal &  & Standard & 0.455 & 0.630 \\
Centerpoint Map & Focal &  & Dynamic & 0.467 & 0.652 \\
Keypoint Heatmap & BCE &  & Standard & 0.457 & 0.641 \\
Keypoint Heatmap & BCE &  & Dynamic & 0.466 & 0.650 \\
Keypoint Heatmap & Focal &  & Standard & 0.487 & 0.668 \\
Keypoint Heatmap & Focal &  & Dynamic & \textbf{0.496} & \textbf{0.681}\\\bottomrule
\end{tabular}
\label{tab:abkey}
\end{table}

\begin{table}[t]
\centering
\caption{Ablation study on backbone and neck selection.}
\begin{tabular}{llcc}
\toprule
Backbone & Neck & mPQ &bPQ \\\midrule
ResNet-101 &FPN & 0.486  & 0.673 \\
ResNext-101-DCN &FPN &0.480   & 0.667 \\
HRNet-w64 & FPN&0.490 &0.675  \\
HRNet-w64 & ASPP  &0.431&0.601\\
HRNet-w64 & Shared JPFM& 0.492 & 0.677 \\
HRNet-w64 & Unshared JPFM& 0.496 & 0.681 \\\bottomrule
\end{tabular}
\label{tab:bneck}
\end{table}

\subsubsection{Backbone And Neck Selection}
\label{sec:neck}
ResNet backbone and FPN~\cite{fpn} are commonly considered the default multi-scale feature extractor in many computer vision tasks due to their simple but efficient architecture. However, as discussed before, a network in series may not be a good selection for local feature extraction. In order to validate this, we performed several experiments using different backbones and neck combinations. As shown in \autoref{tab:bneck}, the combination of ResNet101 and FPN achieved 0.486 and 0.673 in terms of mPQ and bPQ, which was already higher than the previous work HoVer-Net thanks to the benefits of the head design of PointNu-Net. However, when we used ResNeXt-101-DCN~\cite{zhu2019deformable} which is considered more powerful in natural image processing, the performance became even worse. This might be the reason that a better ability in catching global features may not benefit nuclei segmentation tasks. Instead, HRNet-w64 can better utilize local features for classifying and detecting high-density instances, as it keeps the high-resolution features along feature extraction. However, since HRNet-w64 has merged multi-scale features in parallel, the effect of FPN was very limited. Meanwhile, solely using ASPP~\cite{chen2017deeplab} may significantly degrade the performance as it is not designed for multi-scale feature fusion. Hence, we introduced JPFM to make full use of the features from HRNet-w64. For feature fusion, we have two options: to construct a shared JPFM to merge the features for all branches or to learn the feature in each branch separately, which are termed as shared and unshared JPFM respectively. Experimental results demonstrated that a shared JPFM outperformed the default neck FPN, while the unshared JPFM further boosted the network’s ability since different scale information may not have the same importance for the three separated branches. 

\begin{figure}[t]
\centering
\includegraphics[width=0.75\columnwidth]{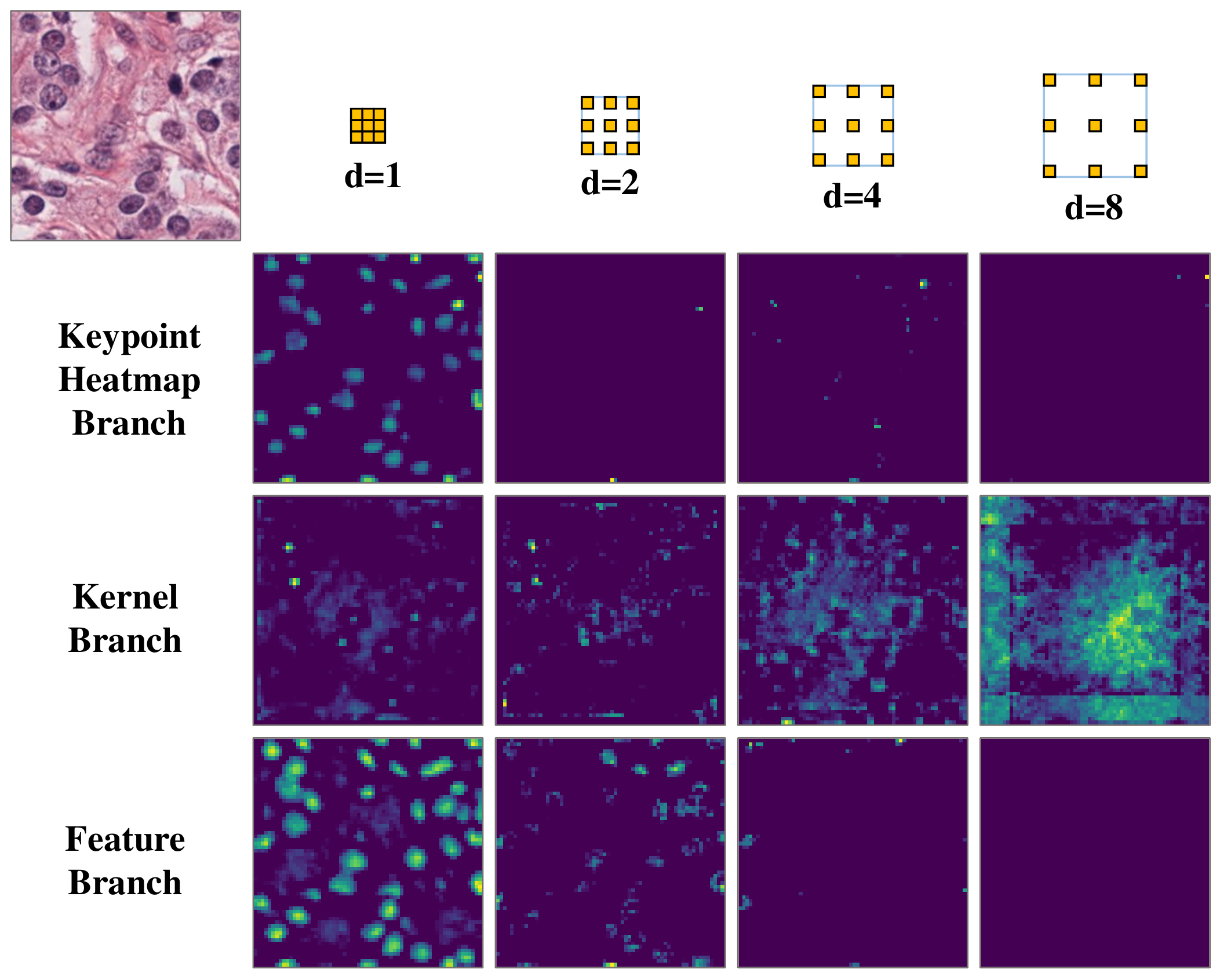}
\caption{\textbf{Visualization of feature activation maps of all convolution layers in the JPFM in the keypoint heatmap branch, kernel branch, and feature branch.} \textbf{d} denotes the dilation of convolution in the JPFM. }
\label{fig:actvisual}
\end{figure}
To better understand the unshared JPFM, we visualized the feature activation map in the three branches, as shown in \autoref{fig:actvisual}. As observed, keypoint heatmap and feature branches had more active feature maps in $d=1$, which means that the low-level high-resolution features are important to detection and classification. In contrast, the feature map in the kernel branch activated strongly in $d=8$, reflecting the crucial role of high-level low-resolution features in the kernel generation.

\section{Conclusion}
In this paper, we have presented a keypoint-aware PointNu-Net for nuclei segmentation and classification within multi-tissue histology images, which detects and classifies nuclei by keypoint heatmap regression and segments nuclei simultaneously via dynamic convolution. 
We have proposed a feature aggregation module to enhance the local and global features from HRNet, called Joint Pyramid Fusion Module, which allows the network to take full use of the merged multi-scale features. In addition, we have conducted the speed-accuracy trade-off experiments and proposed lightweight versions of PointNu-Net for faster inference, which requires no image post-processing and NMS operation. 
Quantitative experiments have shown that our proposed method has achieved state-of-the-art nuclei segmentation, detection, and classification performance.


\section*{Acknowledgments}
The work was partially supported by the following: Jiangsu Science and Technology Programme (Natural Science Foundation of Jiangsu Province) under no. BE2020006-4, and UK Engineering and Physical Sciences Research Council (EPSRC) Grants Ref. EP/M026981/1, EP/T021063/1, EP/T024917/.
\bibliographystyle{IEEEtran.bst}
\bibliography{refs}

\end{document}